\documentclass[aps,preprint]{revtex4}%
\usepackage{amsfonts}
\usepackage{amsmath}
\usepackage{amssymb}
\usepackage{graphicx}%
\begin{document}

\begin{center}
\textbf{To the problem of cross-bridge tension in steady muscle shortening and
lengthening }

.

By Valery B. Kokshenev

.

Submitted to the Journal of Biomechanics 14 April 2009, BM-D\_09-00317 

.

\textit{Departamento de F\'{\i}sica, Universidade Federal de Minas Gerais,
Instituto de Ci\^{e}ncias Exatas, Caixa Postal 702, CEP 30123-970, Belo
Horizonte, Brazil, valery@fisica.ufmg.br}
\end{center}

.

\textbf{Abstract.}  Despite the great success of the Huxley sliding filament
model proposed half a century ago for actin-myosin linkages (cross-bridges),
it fails to explain the force-velocity behavior of stretching skeletal
muscles. Huxley's two-state kinetic equation for cross-bridge proportions is
therefore reconsidered and a new solution to the problem of steady muscle
eccentric and concentric contractions is reported. Instead of numerical
modeling the contractive-force data by appropriate choice of the seemingly
arbitrary heterogeneity of attachment and detachment rates of myosin heads to
actin filament cites, Huxley's idea on mechanical equilibrium is probed into
thermodynamic equilibrium in the whole overlapped actin-myosin zone. When the
second law of statistical thermodynamics is applied to cross-bridge
proportions, the weakly bound states appear to be correlated to the strongly
bound states via structural and kinetic intrinsic muscle characteristics. A
consequent substantial reduction of the number of free parameters in
cross-bridge proportions is also due to the overall self-consistency
(normalization) of attachment-detachment stochastic events. The explicit
force-velocity curve is found to be generic when applied to the reduced
tension in a single cross bridge, sarcomere, fiber, or muscle as a whole
during its active shortening or lengthening. This universal curve fits the
empirical tension-velocity data on frog muscle shortening using only one
adjustable parameter, while the Huxley model employed four parameters. The
established normally distributed cross-bridges, detaching slowly near
equilibrated states in steady lengthening muscle and quickly in shortening
muscle, are in qualitative agreement with recent data on the force enhancement
following muscle stretching.

.

\textbf{1. Introduction}

The early studies of muscle fibers under the light microscope revealed
cross-striations running normal to the fiber axis. It was observed that during
either concentric or eccentric contractions, the length changes occurred via
an increase or decrease in the extent of the I-band with the A-band remaining
unchanged. Two groups laid the foundations for the cross-bridge (CB)\emph{\ }%
theory when they simultaneously suggested that muscle contractions occur due
to the relative sliding of the thick myosin filaments past the thin actin
filaments, mediated by the ATP-dependent actin-myosin linkages (H. E. Huxley
and Hanson, 1954) working as independent force generators (A. F. Huxley and
Niedergerke, 1954).

A. F. Huxley (1957) evaluated muscle tension\ caused by shortening, in fact,
based on the idea of the existence of \emph{mechanical} equilibrium of myosin
heads at the regular sites of actin filaments. His famous two-state
(bound-unbound) sliding filament model was determined on the basis of the
simplest standard kinetic equation controlled by the velocity-independent
rates of\ attachment and detachment of myosin heads. In spite of the great
success in illuminating the force generation and power liberation during
muscle shortening, the Huxley approach generally failed to explain the
ascending branch of the phenomenologically established tension-velocity
equation (see e.g. Harry et al., 1990 and references therein).

Exploring the fact that the sliding filament model leaves a free choice of the
attachment and detachment rate functions, many researchers successfully
simulated a range of muscle properties in lengthening by fitting the empirical
data by linear functions and constants suggested by Huxley (1957) for the
CB\ proportions or by bilinear and exponential functions. Likewise,
considerable efforts have been made to modify the rate functions (Zahalak,
1981; Harry et al., 1990; Ma and Zahalak, 1991; Cole et al., 1996) or to find
an exact \emph{numerical} solution to Huxley's model (Wu et al., 1997). Very
recently, controversies surrounding Huxley's approach were brought forth by
Mehta and Herzog (2008) in their careful studies of force exposed by a single
CB during lengthening.

The theoretical problem of self-consistency in the two-state sliding filament
models was thoughtfully discussed by Hill and co-workers (Hill et al., 1975).
Considering the conditions of CB \emph{thermodynamic} equilibrium besides the
minimum of mechanical energy (Huxley, 1957), they demonstrated that the
original Huxley model has a low efficiency in comparison to its modified
versions. Moreover, it was noted by Eisenberg et al. (1980) that "there is no
in vitro evidence for ... the basic (Huxley's) assumption that the
cross-bridge detaches slowly". The lacking data were provided by Mehta and
Herzog (2008).

In this study, I develop a statistical approach to the filament sliding
mechanism and show that only linear functions for the attachment-detachment
rates are compatible with the concept of thermodynamic equilibrium. A new
analytical solution to Huxley's two-state kinetic equation is proposed and
verified using the available from the literature data on tension in steady
muscle lengthening and shortening.

.

\textbf{2. Methods}

\textit{2.1. Model by Huxley (1957) revisited}

At a fixed muscle \emph{contraction\ velocity }$V$, the number of bound CB
states $N_{V}$ combining myosin filament with actin filament of the total
number of sites $N_{0A}$ obeys the common "balance" kinetic equation%
\begin{equation}
\frac{d}{dt}N_{V}(x,t)=\frac{\partial N_{V}}{\partial t}+\frac{\partial N_{V}%
}{\partial x}\cdot\frac{dx}{dt}=\text{ }f(x)(N_{0A}-N_{V})-g(x)N_{V}.
\label{A1}%
\end{equation}
Here $f$ and $g$ are \emph{attachment} and \emph{detachment rates} of the
corresponding unbound and bound states located in time $t$ at a distance $x$
estimated from the nearest site $x=0$. The steady process determined by late
times $t\gg f^{-1},$ $g^{-1}$ providing $\partial N_{V}(x,\infty)/\partial
t=0$ in Eq. (\ref{A1}), reduces Eq. (\ref{A1}) to%
\begin{equation}
-\frac{V}{2}\frac{d}{dx}n_{V}(x)=\text{ }f(x)(1-n_{V})-g(x)n_{V}\text{,}
\label{sta eq}%
\end{equation}
\textrm{ }i.e., to Huxley's Eq. (4)\textrm{ }where the \emph{proportion}
$n_{V}(x)=N_{V}(x,\infty)/N_{0A}$ of CBs during \emph{steady shortening}.
According to Huxley, the force output $F(x)=kx$ is produced when $x$ decreases
at a positive velocity of sliding of the actin filament $V_{A}$ and a negative
velocity of myosin filament $V_{M}$, i.e., $V_{A}=-V_{M}=-dx/dt>0$. The
contraction velocity per one-half sarcomere $V/2$ determines the contraction
\emph{velocity }$V$\emph{\ }of the muscle\emph{ }as a whole, when modeled by
$V=V_{A}-V_{M}=2V_{A}$. As can be derived from Huxley's Eq. (6) with the
preservation in part its notations, the overall generated force
\begin{equation}
F_{V}^{(total)}=\lim_{L\rightarrow\infty}\frac{sN_{0M}}{L}\int_{-L}%
^{+L}F(x)n_{V}(x)\frac{dx}{2l_{A}}\text{,} \label{A2}%
\end{equation}
was evaluated via the force $F(x)$ per one \emph{myosin site}, as one actin
site is carried past it. Here $s$ is the sarcomere length, $l_{A}$ is the
\emph{trial} distance between the nearest sites in the actin filament
\emph{evaluated} in Huxley's Eq. (15), and $N_{0M}$ is the number of sites in
the thick filament in the overlapping zone of length $L$.

The solution to Eq. (\ref{sta eq}) for the shortening regime (hereafter
distinguished by index $1$) was found as a combination of the \emph{localized}
(short-domain) and \emph{delocalized} (large-domain) spatially correlated
(bound) states. The corresponding proportions reproduced exactly from Huxley's
Eqs. (7) and (8) are%
\begin{equation}
n_{V}^{(loc)}(x)=n_{01}\left(  1-\exp\left[  \frac{V_{1}}{V}\left(
\frac{x^{2}}{h^{2}}-1\right)  \right]  \right)  \text{, }0\leq x\leq h
\label{A4}%
\end{equation}
and%
\begin{equation}
n_{V}^{(deloc)}(x)=n_{01}\left[  1-\exp\left(  -\frac{V_{1}}{V}\right)
\right]  \exp\left(  2x\frac{g_{1}^{\prime}}{V}\right)  \text{, }-\infty
<x\leq0\text{,} \label{A5}%
\end{equation}
though parameterized here by%
\begin{equation}
n_{01}=\frac{f_{1}}{f_{1}+g_{1}}\text{ and }V_{1}=h(f_{1}+g_{1})\text{.}
\label{A6}%
\end{equation}
In turn, this description of the two CB\ states follows from the rates
\emph{postulated} by linear functions, namely%
\begin{equation}
f(x)=f_{1}\frac{x}{h}\text{ and }g(x)=g_{1}\frac{x}{h}\text{, for }0\leq x\leq
h\text{,} \label{A3}%
\end{equation}
and two constants $f^{\prime}(x)=0$, $g^{\prime}(x)=g_{1}^{\prime}$, for
$-\infty<x<0$. The muscle concentric steady tension $P_{V}$ reduced to the
model isometric tension $P_{0}$ found on the basis of Eqs. (\ref{A2}%
)-(\ref{A3}), namely%
\begin{equation}
\frac{P_{V}^{(short)}}{P_{0}}=1-\frac{V}{V_{1}}\left[  1-\exp\left(
-\frac{V_{1}}{V}\right)  \right]  \left(  1+\frac{VV_{1}}{2h^{2}g_{1}%
^{\prime2}}\right)  \text{, for }V>0\text{,} \label{A7}%
\end{equation}
was fitted by the widely cited four model parameters: $f_{1}=43.3$ $s^{-1}$
and $g_{1}=10.0$ $s^{-1}$, indicating slow detachment of the localized CBs,
and $f_{1}^{\prime}=0$ with $g_{1}^{\prime}=209$ $s^{-1}$, for delocalized
states. In addition, two more adjustable parameters $h\thickapprox15$ $nm$ and
$V_{1}=V_{\max}^{(\exp)}/4$, where $V_{\max}^{(\exp)}$ is the empirical
maximum shortening velocity, were indirectly employed when tested by Hill's
empirical equation (see Chapter IV in Huxley, 1957). It is noteworthy that the
nearest-site distance in the actin filament treated as a free parameter was
estimated as $l\thickapprox h$, i.e. close to the known nearest-molecular
distance in the myosin filament $l_{M}=14.5$ $nm$ (Craig and Woodhead, 2006).
However, the ratio $P_{-\infty}/P_{0}=(f_{1}+g_{1})/g_{1}=5.33$ reported by
Huxley (1957) for the muscle lengthening regime, contrasts to the observed
ratios falling between $1.8$ and $2.0$ (e.g. Harry et al., 1990).

.

\textit{2.2. A new solution to Huxley's steady equation}

Beyond any specific suggestions, the formal solution to the steady-state Eq.
(\ref{sta eq})%
\begin{equation}
n_{V}(x)=n_{0}(x)+\Delta n_{V}(x)=n_{0}+(1-n_{0})c_{V}\exp\left(  -\frac
{1}{\overset{\cdot}{x}}\int_{0}^{x}[f(x^{\prime})+g(x^{\prime})]dx^{\prime
}\right)  ,\text{ }\overset{\cdot}{x}\equiv\frac{dx}{dt}=\mp\frac{V}%
{2}\text{,} \label{nx}%
\end{equation}
is valid for any contraction shortening velocity $V$ ($=-$ $2\overset{\cdot
}{x}>0$) and lengthening velocity $V$ ($=2\overset{\cdot}{x}<0$), leaving an
arbitrary choice of the rate functions $f(x)$ and $g(x)$. A differential
equation of the first order possesses as common only one free constant,
denoted by $c_{V}$, whereas%
\begin{equation}
n_{0}(x)=\frac{f(x)}{f(x)+g(x)} \label{n0}%
\end{equation}
straightforwardly following from Eq. (\ref{sta eq}) taken at $V=0$, describes
maximal CB proportions limited by intrinsic rates. In the Huxley model, the
constant $c_{V}=-n_{01}(1-n_{01})^{-1}\exp(-V_{1}/V)$ in Eq. (\ref{nx})
results from his \emph{boundary condition} $n_{V}(h)=0$ providing the
\emph{non-Gaussian} proportion (\ref{A4}) for CB\ localized states during
muscle shortening.

Besides the boundary conditions considered below, let us employ the property
of periodicity in the overlapping part of the actin filament\textrm{ }of
length $Nd$ having $N$ \emph{occupied} \emph{cells}. Since the Huxley
proportion $n_{V}(x)$ in Eq. (\ref{nx}) plays the role of the late-time
\emph{probability} of finding one of the two myosin heads attached at a
position $x$ between two nearest equivalent sites (see also Hill et al. 1975,
p. 346), the total force output in a \emph{finite} overlapped zone is%
\begin{equation}
F_{V}^{(zone)}=NF_{V}=\int_{-Nd}^{+Nd}F(x^{\prime})n_{V}(x^{\prime}%
)\frac{dx^{\prime}}{2d}=N\int_{-d}^{d}F(x)n_{V}(x)\frac{dx}{2d}\text{, where
}x^{\prime}=xN\text{.} \label{Ftot}%
\end{equation}
Here $F(x)$ is the active force \emph{per one} \emph{actin site}, substituting
that per one myosin cite in Eq. (\ref{A2}). Such a consideration suggests the
statistical equivalence of all the occupied cells in the actin filament
treated as a one-dimensional crystal of lattice constant $d$ ($=36$ $nm$, e.g.
Hill et al., 1975). In order to be consistent with Eqs. (\ref{Ftot}) and
(\ref{A2}), the normalization conditions for the CB distributions
(proportions)%
\begin{equation}
\int_{-d}^{+d}n_{V}(x)\frac{dx}{2d}=\int_{0}^{d}n_{V}(x)\frac{dx}{d}=\int
_{-d}^{0}n_{V}(x)\frac{dx}{d}=1\text{ } \label{n-str-sh}%
\end{equation}
must be taken into consideration.

Furthermore, extending Huxley's idea on the minimum of mechanical energy at
$x=0$ over the minimum of Gibbs energy (Eisenberg et al., 1980), the CB state
with $F(0)=0$ is treated as the locally equilibrated state, having
\emph{maximum configurational entropy} at $x=0$. As the consequence of one of
the most general principle (second law) of statistical thermodynamics, the
distribution of bound states $n_{V}(x)$ given in Eq. (\ref{nx}) must have
Gaussian form centered at $x=0$ (see e.g. Chapter 12 in Landau and Lifshitz,
1989). One can see from Eq. (\ref{nx}), that the thermodynamical principle
ensured by the \emph{sign requirement} $\overset{\cdot}{x}x>0$ can be
satisfied solely by the linear parameterization of the rate functions, namely%

\begin{align}
f(x)  &  =f_{m}\frac{x}{x_{m}}\text{, }g(x)=g_{m}\frac{x}{x_{m}}%
\text{,}\nonumber\\
\text{for }x_{m}  &  =x_{+}\geq x\geq0\text{ or }x_{m}=-x_{-}\leq
x\leq0\text{,} \label{fx-gx}%
\end{align}
Consequently, the normalization constant%
\begin{equation}
c_{V}=\frac{d}{|x_{m}|}\frac{2}{\sqrt{\pi v}}\operatorname{erf}\left(
\frac{1}{\sqrt{v}}\right)  ^{-1}\text{, }v=\frac{V}{V_{m}}=\frac
{2\overset{\cdot}{x}}{(f_{m}+g_{m})x_{m}}>0\text{,} \label{cv}%
\end{equation}
readily follows from the normalization conditions (\ref{n-str-sh}), where the
standard \emph{error function }$\operatorname{erf}(y)=(2/\sqrt{\pi})\int
_{0}^{y}\exp(-t^{2})dt$, lying between $0$ [$=\operatorname{erf}(0)$] and $1$
[$=\operatorname{erf}(\infty)$], is employed.

.

\textbf{3. Results}

\textit{3.1. CB proportions in steady muscle shortening and lengthening}

As seen in Eq. (\ref{nx}), a given CB is characterized by the equilibrated
velocity-independent \emph{ground} state and the \emph{excited}
non-equilibrated state described, respectively, by uniform proportion
$n_{0}=f_{m}/(f_{m}+g_{m})$ and non-uniform, heterogenous stochastic
proportion $\Delta n_{V}(x)$, having the meaning of probabilities normalized
in Eq. (\ref{n-str-sh}). The requirement of signs ( $\overset{\cdot}{x}x>0$),
while in particular ensuring the self-consistency with the ground state (when
$x\rightarrow0$, $\Delta n_{V}(x)\rightarrow0$), also constrains possible
domains for both CB states, as shown in Eq. (\ref{fx-gx}). Indeed, both kinds
of domains ($0\leq x\leq x_{0}$ and $-x_{0}\leq x\leq0$, otherwise
$n_{0}(x)=0$) are generally possible for the ground state, whereas only the
negative domain, $-x_{-}\leq x\leq0$, satisfies the requirement of
self-consistency of the solution given in Eq. (\ref{nx}). In this way, the CB
boundary conditions are not postulated as in Eqs. (\ref{A4})-(\ref{A3}), but
result from the minimum of Gibbs energy, also giving rise to the CB
\emph{mechanical constraints}, consistent with the simultaneous observations
of directions of both the force output and contraction velocity, as
illustrated in Fig. 1.

.

\textbf{Place Fig. 1}

.

The analysis in Fig. 1 specifies domains of the \emph{short-domain} CB states
which, being incorporated in the trial Eq. (\ref{nx}) with the help of Eq.
(\ref{cv}), yield%
\begin{equation}
n_{V}(x)=n_{0}\Theta_{0}(x)+(1-n_{0}\frac{x_{0}}{d})\frac{d}{x_{-}}\frac
{2}{\sqrt{\pi v}}\frac{\exp\left(  -\frac{x^{2}}{vx_{m}^{2}}\right)
}{\operatorname{erf}\left(  \frac{1}{\sqrt{v}}\right)  }\Theta_{V}(x)\text{.}
\label{nV}%
\end{equation}
Here, the auxiliary functions $\Theta_{0}(x)=\Theta(\pm x)-\Theta(x\mp x_{0})$
and $\Theta_{V}(x)\equiv\Theta(-x)-\Theta(x+x_{-})$ are introduced by the
standard Heaviside (step) function $\Theta(y)$, which is one for $y\geq0$ and
zero for $y<0$.

.

\textit{3.2. Muscle tension in steady shortening and lengthening}

The mean force output $F_{V}$ generated by a single cell of the actin filament
is evaluated using Eqs. (\ref{Ftot}) and (\ref{nV}), namely%
\begin{equation}
F_{V}=F_{0}+\Delta F_{V}=k\int_{-d}^{+d}xn_{V}(x)\frac{dx}{2d}=F_{0}%
-k\frac{x_{-}}{2}(1-n_{0}\frac{x_{0}}{d})\Phi(v)\text{, }F_{0}=\pm\frac
{kx_{0}^{2}}{2d}n_{0}\text{,} \label{FV}%
\end{equation}
via the CB \emph{stiffness} $k=F(x)/x$ (Huxley, 1957; Huxley and Simmons,
1971), shown to be a velocity-independent intrinsic muscle quantity (e.g.
Lombardi and Piazzesi, 1990), where
\begin{equation}
\Phi(v)=2\sqrt{\frac{v}{\pi}}\frac{1-\exp(-\frac{1}{v})}{\operatorname{erf}%
(\frac{1}{\sqrt{v}})}\text{; with }\Phi(0)=0\text{, }\Phi(1)=0.846\text{, and
}\Phi(\infty)=1\text{.} \label{Fi}%
\end{equation}
The upper and lower signs in the velocity-independent limiting steady force
$F_{0}$ (\ref{FV}) correspond to shortening and lengthening (see Fig. 1). In
this way, Huxley's Eq. (\ref{A7}) is transformed into a unique equation%
\begin{equation}
\frac{P_{V}}{P_{0}}=\frac{F_{V}}{F_{0}}=1\mp\sigma_{m}\Phi\left(  v\right)
\text{, with }\sigma_{m}=\frac{(d-n_{0}x_{0})x_{-}}{n_{0}x_{0}^{2}}\text{, }
\label{PV}%
\end{equation}
for the reduced CB tension and force output in both concentric and eccentric
muscle contractions conducted at positive and negative steady velocities
$V=vV_{m}$ (\ref{cv}), respectively.

The \emph{one-parameter}\ fitting analysis of the proposed theory is conducted
on the basis of Eq. (\ref{PV}) and the available experimental data. In Fig. 2,
the muscle shortening is described by%
\begin{align}
\text{ }\frac{P_{V}^{(short)}}{P_{01}}  &  =1-\frac{\Phi\left(  \frac{\lambda
V}{V_{\max}}\right)  }{\Phi(\lambda)}\text{, }0\leq V\leq V_{\max}\text{,
}\nonumber\\
V_{\max}  &  =\lambda V_{m1}\text{, }V_{m1}=x_{m1}(f_{m1}+g_{m1})\text{,}
\label{psh}%
\end{align}
where $\lambda$ is an adjustable parameter.

.

\textbf{Place Fig. 2}

.

.

\textbf{Place Fig. 3}

.

The steady muscle lengthening is fitted in Fig. 3 by%
\begin{align}
\frac{F_{V}^{(stret)}}{F_{02}}  &  =1+\Phi\left(  \frac{V}{V_{m2}}\right)
\text{, }-\infty<V\leq0\text{, }\nonumber\\
V_{m2}  &  =-x_{m2}(f_{m2}+g_{m2})<0\text{.} \label{pst}%
\end{align}
using the characteristic velocity $V_{m2}$ as a free parameter. Other
parameters describing two distinct regimes are specified as $x_{-}=x_{m1}$,
$f_{m}=f_{m1}$, $g_{m}=g_{m1}$, for shortening, and $x_{-}=x_{m2}$,
$f_{m}=f_{m2}$, $g_{m}=g_{m2}$, for lengthening. One can see that $V_{\max}$
plays the role of the maximum shortening velocity at which $P_{V}^{(short)}%
=0$, and $V_{m2}$ is a characteristic velocity separating slow and fast
lengthening. Also, the limiting tension in the fastest steady lengthening is
$P_{-\infty}^{(stret)}=2P_{0}$.

.

\textit{3.3. CB domains}

The force-velocity fitting analysis alone does not provide details on the CB
attachment-detachment rates or the domains. Physically, these domains follow
from the conditions of realization of thermodynamic stability described by a
minimum of Gibbs energy (Hill et al., 1975). Nevertheless, the overall curve
conditions of observation can be established here by the inequalities
$\sigma_{m1}^{(\exp)}>1\geq\sigma_{m2}^{(\exp)}$, resulting from Eqs.
(\ref{psh}) and (\ref{pst}), where the fitting parameter $\sigma_{m1}^{(\exp
)}=\Phi(0.85)^{-1}=1.22$ is found for muscle shortening and $\sigma
_{m2}^{(\exp)}$, generally lying between $0.8$ and $1.0$, for lengthening.

Alternatively, the observation conditions of the predicted branches of the
master curve can be reformulated in terms of the CB rigor state proportions
$n_{01}<x_{m1}dx_{01}^{-1}(x_{01}+x_{m1})^{-1}$ and $n_{02}\geq x_{m2}%
dx_{02}^{-1}(x_{02}+x_{m2})^{-1}$, obtained with the help of Eq. (\ref{PV}).
This finding can be improved when the CB geometrical constraints shown in Fig.
1 are taken into account. Indeed, since the tail of the myosin molecule is
longer than heads, one should expect a geometrical constraint $x_{m2}>x_{m1}$,
providing $n_{02}>n_{01}$. Under the simplified requirements of periodicity
($x_{01}+x_{m1}=d$ and $x_{02}=x_{m2}=d$), the CB proportions underlying the
observation of the master curve are specified in the insets in Figs. 2 and 3.

.

\textbf{4.} \textbf{Discussion}

Huxley's model of the establishment of mechanical equilibrium of myosin heads
near actin-filament sites is based on the simplest kinetic equation
determining a balance between unbound and bound actin-myosin states. In a
muscle contracting at constant velocity $V$, these two states are described by
the proportions $1-n_{V}(x)$ and $n_{V}(x)$, satisfying the steady-state
kinetic equation (\ref{sta eq}) at generally arbitrary rates $f(x)$ and
$g(x)$. Such a property, following evidently from the solution $n_{V}(x)$
found for a general case in Eq. (\ref{nx}), implies that the kinetic equation
accounts for the most general features of muscle relaxation, regardless of
details underlying the attachment-detachment mechanism of myosin heads.
Consequently, theoretical studies exploring an arbitrary choice of the
functional form of the attachment-detachment rates, which involved increasing
number of numerical parameters, guarantee good fit to phenomenological data,
but do not shed light on the muscle intrinsic characteristics.

After work by Rayment et al. (1993) on the structural study of force
generators in contracting muscles, the observations of catalytic domains of
myosin being initially weakly attached to actin are commonly associated with
the \emph{weakly bound }CB states, and the following structural changes
resulting in tight binding of actin-myosin linkages are associated with
\emph{strongly bound }CB states. Since the steady-state equation
(\ref{sta eq}) is a late-time part of more general kinetic equation
(\ref{A1}), the proportions $n_{V}(x)$ are also part of the non-steady
solutions, as demonstrated by Lombardi and Piazzesi (1990) and recently by
Walcott and Herzog (2008) employing Huxley's Eqs. (\ref{A4}) and (\ref{A5}).
It seems therefore plausible to associate Huxley's short-domain proportion
$n_{V}^{(loc)}(x)$ and large-domain proportion $n_{V}^{(deloc)}(x)$ with
respectively weak and strong late-time CB states. In this study, the
actin-myosin bound state is composed of the equilibrated and excited states
distributed by Gaussian function dictated by the second law of thermodynamics.

Within the proposed framework of stochastic approach to the
attachment-detachment events of myosin heads, a common requirement of
normalization of the random proportion $n_{V}(x)$ specifies the heterogeneity
of the CB distribution via the correlated intrinsic structural ($x_{m}$, $d$),
kinetic ($f_{m}$, $g_{m}$) and dynamic ($V_{m}$) muscle characteristics [see
e.g. Eq. (\ref{cv})], that decreases the number of free parameters. Moreover,
the trend of weakly bound myosin heads to achieve maximum structural-domain
entropy in the vicinity of actin-filament sites ($x\approx0$) requires a
correlation in signs between the head displacements ($x$) and velocities
($\overset{\cdot}{x}$). Consequently, the conceivable CB\ domains for both
bound states schematically shown in Fig. 1 are eventually described by
Heaviside functions in Eq. (\ref{nV}). A geometrical selection of the main
components of the\textrm{ }force resulting in the power stroke in a direction
consistent with the vector of contraction velocity are also shown in Fig.1.

The explicit solution (\ref{nx}) to Huxley's kinetic equation for CB
proportions $n_{0}(x)$ and $\Delta n_{V}(x)$, distributing respectively
strongly and weakly bound states over the actin filament cells, results in the
velocity-independent (isometric) force $F_{0}$ and contractive force $\Delta
F_{V}$, components of the CB force output $F_{V}$ (\ref{FV}). In Fig. 2,
famous Huxley's comparative analysis with Hill's data on muscle concentric
tension (Huxley, 1957, p. 287) is revisited. The high-velocity wing of the
tension curve above $V/V_{\max}=0.5$ controlled mostly by weak CB\ states is
well fitted by both non-Gaussian (\ref{A5}) and Gaussian (\ref{nV})
proportions. It is not the case of the low velocity region $0.2<V/V_{\max
}<0.5$, where a discrepancy between Huxley's curve (\ref{A7}) and the data
indicate a disadvantage of the short-domain ($x<h<d$) weak CBs exerting
negative force $\Delta F_{V}$ and eventually reducing the total produced
tension. In contrast to the postulated retarded detachment ($g_{1}<f_{1}$)
discussed in Eqs. (\ref{A4}) and (\ref{A3}), the Gaussian strong and weak CBs
require faster detachment than attachment ($g_{m1}>f_{m1}$) in muscle
shortening, as derived from Hill's data and shown in the inset in Fig. 2. The
regular deviation of Gaussian CBs from the data at very low velocities is
associated with a simplified modeling of the channel of relaxation of weak
states to strong states. Indeed, the fit analyses can be improved when the
proportion of\ a new weak-to-strong transient Gaussian CB spreads its domain
symmetrically within the range $-\delta\leq x\leq\delta$, extending the weak
CB state in the vicinity of $x\approx0$, as shown by the dotted line in the
inset in Fig. 2 for the case $V/V_{\max}=0.1$.

In Fig. 3, the upper branch of the force-velocity curve (\ref{PV}) drawn at a
single adjustable parameter ($V_{m2}=-288$ $nm/s$) fits well the empirical
data on CB force in muscle stretching. Similar to shortening, the fitting
analysis could be improved at low stretch velocities when a transient bound
state is additionally introduced, as independently proposed by Mehta and
Herzog (2008, Fig. 3). These authors also raised the central question on the
existence of Huxley's proportions favoring myosin head attachment events at
large distances with an increase in contraction velocity. One therefore infers
that although non-Gaussian large-domain CBs (\ref{A5}) numerically fit the
empirical data (Fig. 2), they do favor neither thermodynamic equilibrium in
the overlapped zone nor high cycle efficiency (Hill et al., 1975).

The microscopic structures of the short-domain bound states are provided above
via observation conditions of the generic curve (\ref{PV}) equally applied to
the reduced tension in a single CB, sarcomere, fiber, or muscle as a whole
during its steady shortening or lengthening. It is also demonstrated (inset in
Fig. 3) how the two-state muscle cycle \emph{duty ratio} $\beta$ derived from
real experiments can be helpful in a characterization of the Gaussian CB
\emph{rate ratio} $\alpha=f_{m}/g_{m}$ [$=(1-\beta)/\beta$] and strong bound
state proportion $n_{0}$ ($=1-\beta$). To summarize a comparative analysis of
structural and kinetic characteristics of CBs, one can see that \emph{steady}
muscle eccentric and concentric contractions are well distinguished via the
attachment-detachment rate rations, with $\alpha_{stret}>1>\alpha_{short}$,
the strongly-bound occupation numbers, with$\ n_{0}^{(stret)}>n_{0}^{(short)}%
$, supported by the directly observable cycle duty ratios $\beta_{stret}%
<\beta_{short}$ (Mehta and Herzog, 2008). Within this context, the working
hypothesis by Mehta and Herzog (2008) "that a stretched cross-bridge might
remain attached longer than a cross-bridge that had been shortened while
attached" combined with the main finding by Lombardi and Piazzesi (1990) that
"reattachment (in steady lengthening is)... faster than attachment in the
isometric condition or during shortening ... in the same domain of $x$"
results in the predictions $\alpha_{stret}>1$, $n_{0}^{(stret)}>1/2$, and
$\beta_{stret}<1/2$, which are generally consistent with the CB parameters
derived in the insets in Figs. 2 and 3.

To conclude, the provided statistical thermodynamic analysis of the
attachment-detachment CB\ process, modifying Huxley's mechanical sliding
filament model, can also be figured out as an two-headed steady walking of
synchronous myosin molecules over periodical sites of actin filaments with
multiple $36$-$nm$ steps, as directly observed by Sakamoto et al. (2008). The
fluctuating steps are statistically scattered by the normal distribution,
having the zero mean and variance linear with muscle contraction velocity. The
proposed steady contraction dynamics is universally observable through the two
branches of the force-velocity curve generic for steady shortening and
lengthening of a muscle as a whole or its counterparts. The microscopic
structural muscle characteristics appear to be strongly correlated to kinetic
and dynamic characteristics distinguished by the force output directions
generated in distinct muscle regimes. At a macroscopic level, similar kind of
correlations driven by maximum generated force were revealed via the primary
muscle functions well distinguished though the muscle structure adapted to
efficient eccentric, isometric, or concentric contractions (Kokshenev, 2008).

.

\textbf{Acknowledgements}

.

The author thanks Scott Medler for helpful comments. The financial support by
CNPq is also acknowledged.

.

\textbf{References}

Cole, G. K., Bogert, A. J., Herzog W., Gerritsen, K. G. M., 1996. Modelling of
force production in skeletal muscle undergoing stretch. Journal of
Biomechanics, 29, 1091-1104.

Craig, R., Woodhead, J. L., 2006. Structure and function of myosin filaments.
Current Opinion in Structural Biology, 16, 204--212.

Eisenberg, E., Hill, T.L., Chen, Y.D., 1980. Cross-bridge model of muscle
contraction. Quantitative analysis, Biophysical Journal, 29, 195-227.

Harry, J. D., Ward, A. W., Heglund, N.C., Morgan, D. L., McMahon, T. A., 1990.
Cross-bridge cycling theories cannot explain high-velocity lengthening
behavior in frog muscle, Biophysical Journal, 57, 201-208.

Hill, A. V., 1938. The heat of shortening and the dynamic constants of muscle.
Proceedings of Royal Society of London, 126, 136-195.

Hill, T.L., Eisenberg, E., Chen, Y., Podolsky, R. J., 1975. Some
self-consistent two-state sliding filament models of muscle contraction.
Biophysical Journal, 15,\ 335-372.

Huxley, A. F., 1957. Muscle structure and theories of contraction. Progress in
Biophys. Biophys. Chemistry, 7, 255-318.

Huxley, H. E., Hanson, J., 1954. Changes in the cross-striations of muscle
during contraction and stretch and their structural interpretation. Nature,
173, 973-976.

Huxley, A. F., Niedergerke, R., 1954. Interefrenec microscopy of living muscle
fibers. Nature 173, 971-973.

Huxley, A. F., Simmons, R. M., 1971. Proposed mechanism of force generation in
striated muscle. Nature 233, 533-538.

Huxley, A.F., 1998. Biological motors: Energy storage in myosin molsecules.
Current Biology, 8, R485-R488.

Kokshenev, V. B., 2008. A force-similarity model of the activated muscle is
able to predict primary locomotor functions. Journal of Biomechanics, 41, 912--915.

Landau L. D., Lifshitz, E. M., 1989. Statistical Physics, Pergamon Press, London.

Lombardi, V., Piazzesi, G., 1990. The contractile response during steady
lengthening of stimulated frog muscle fibres. The Journal of Physiology 431, 141-171.

Ma, S., Zahalak, G. I., 1991. A distribution-moment model of energetics in
skeletal muscle. Journal of Biomechanics, 24, 21-35.

Mehta, A., Herzog, W., 2008. Cross-bridge induced force enhancement? Journal
of Biomechanics, 41, 1611-1615.

Rayment, I., Holden, H. M., Whittaker, M., Yohn, C.B., Lorenz, M., Holmes,
K.C., Milligan, R.A., 1993. Structure of the actin-myosin complex and its
implications for muscle contraction. Science, 261, 58-65.

Sakamoto, T., Webb, M. R., Forgac, E., Howard, D. White, H. D., Seller, J.R.
2008. Direct observation of the mechanochemical coupling in myosin Va during
processive movement. Nature 455, 128-132.

Walcott, S., Herzog W., Modeling residual force enhancement with generic
cross-bridge models, 2008. Mathematical Biosciences, 216, 172--186.

Wu, J.Z., Herzog W., Cole G. K., 1997. Modeling dynamic contraction of muscle
using the cross-bridge theory, Mathematical Biosciences, 139, 69-78.

Zahalak, G. I., 1981. A distribution-moment approximation for kinetic theories
of muscular contraction. Mathematical Biosciences, 55, 89-114.

\newpage

\textbf{Figure Legends}

\includegraphics[width=\hsize]{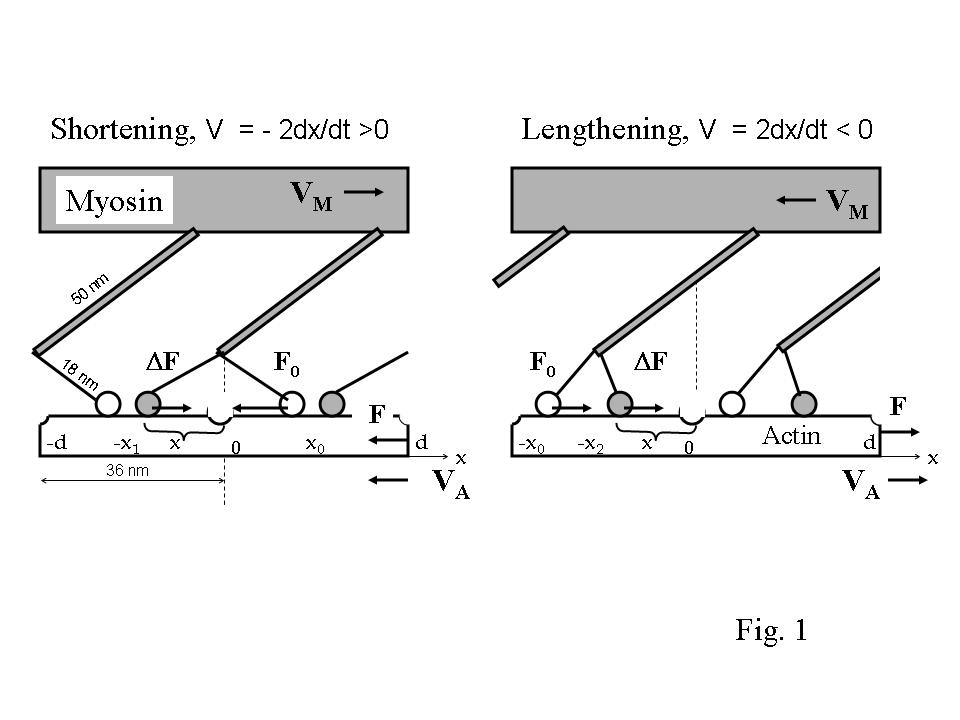}

\textbf{Figure 1.} Mechanical scheme of the force generation by combining
myosin heads with periodic actin filament. Each of the two heads of the
effective CB may be attached to actin filament either in equilibrated ground
state (shown by the open circle) with the uniform probability $n_{0}$, within
the domains $x\leq\pm x_{0}$, or in the non-equilibrium, excited state (closed
circle) with the probability $\Delta n_{V}(x)$, within the domains
$x\leq-x_{1},-x_{2}$. The \emph{arrows} indicate the directions of the sliding
velocity of the actin filament $V_{A}$ and the myosin filament $V_{M}$. During
concentric muscle contraction with a \emph{positive} velocity $V$, the
velocity-independent portion of the generated force $F_{0}$ is also positive,
whereas the ATP hydrolysis results in the negative portion of the contractive
force $\Delta F_{V}$ . During eccentric contractions commonly associated with
the \emph{negative} direction of velocity $V$, both the forces are also negative.

\newpage 

\includegraphics[width=\hsize]{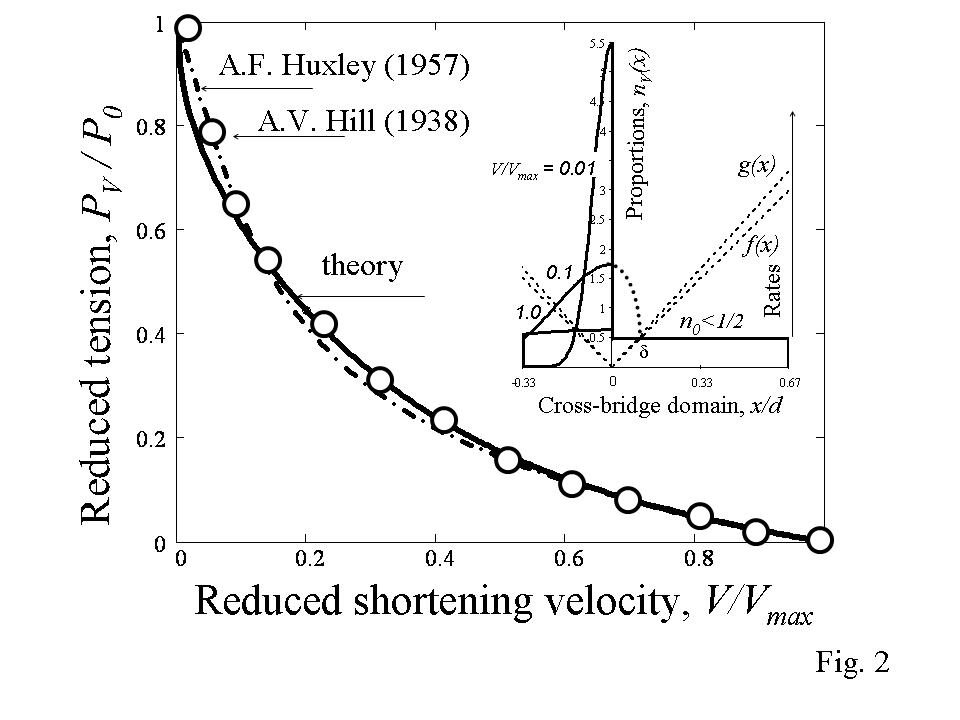}

\textbf{Figure 2}. Analysis of the theoretically predicted tension-velocity
curve using available data on the reduced tension during steady muscle
shortening. The \emph{points} and \emph{dashed-point curve} are the famous
data by Hill (1938) for isolated frog muscles modeled by Huxley (1957), drawn
respectively by the phenomenological equation $P^{(\exp)}/P_{0}=a(1-V/V_{\max
})/(a+V/V_{\max})$, with $a=0.25$ and Eq. (8), fitted by Huxley's parameters
listed above. The \emph{solid line} is Eq. (19) taken at $\lambda=0.85$.
\emph{Inset:} The attachment-detachment rates and CB proportions predicted in
Eq. (15) within the CB\ domains at distinct shortening velocities reduced to
the maximum velocity. The CB structure discussed in the Results is exemplified
by the model parameters $x_{01}^{(\operatorname{mod})}=2d/3$ and
$x_{m1}^{(\operatorname{mod})}=-d/3$, as well as by $n_{01}%
^{(\operatorname{mod})}=0.47$, providing the \emph{rate ratio} $\alpha
_{1}=f_{m1}/g_{m1}=0.88$. The \emph{dotted line} shows a proportion for the
modeled transient CB state schematically drawn for $V/V_{\max}=0.1$.

\newpage
\includegraphics[width=\hsize]{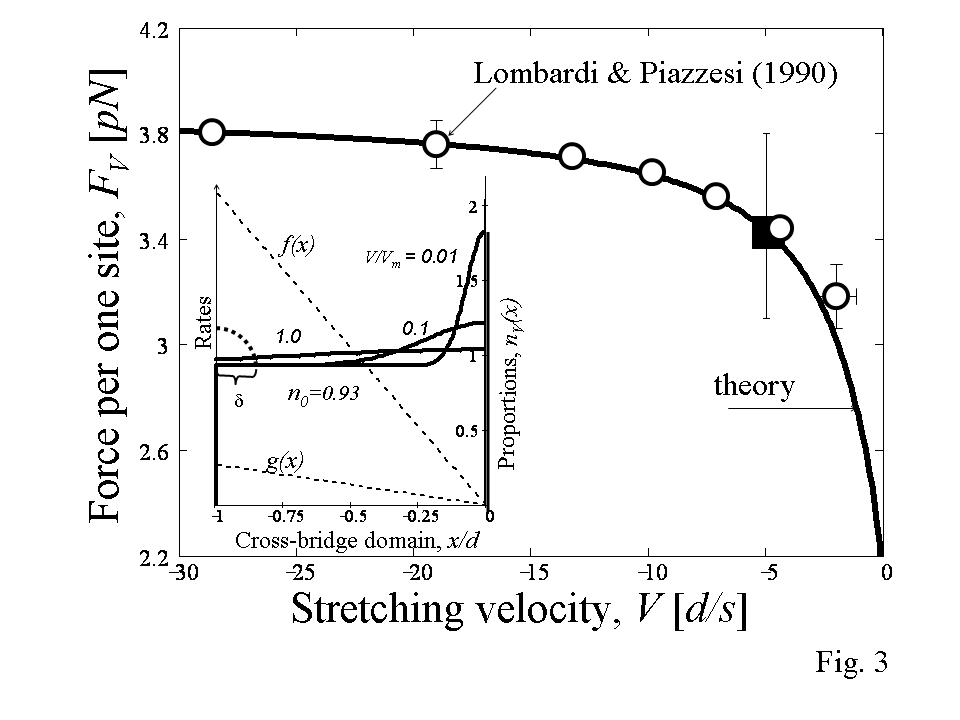}

\textbf{Figure 3}. Steady force induced by one cross-bridge versus the
stretching velocity. The \emph{open circles} are the mean datapoints of the
forces (re-scaled by $|F_{0}|=1.95$ $pN$ ) measured by Lombardi and Piazzesi
(1990, Fig. 7) in frog muscle fibers at stretch velocities lying between $75$
and $1030$ $nm/s$ and scaled here by $d=36$ $nm$. The \emph{closed square}
indicates the force per one CB\ reported by Mehta and Herzog (2008) for
unspecified velocities. The theoretical curve is drawn based on Eq. (20) with
$V_{m2}^{(\operatorname{mod})}=-8$ $d/s$. \emph{Inset:} The
attachment-detachment rates within the CB\ domains and CB proportions
predicted in Eq. (15) at three distinct velocities reduced to the found
$V_{m2}^{(\operatorname{mod})}$. They are exemplified by model parameters
$x_{02}^{(\operatorname{mod})}=$ $x_{m2}^{(\operatorname{mod})}=-d$,
consistent with the observation conditions discussed in the Results, as well
as by $n_{02}^{(\exp)}=1-\beta_{2}^{(\exp)}=0.93$, where the stretch
\emph{cycle} \emph{duty ratio} $\beta_{2}^{(\exp)}=7.35\%$ \ (the time of
attachment $f_{m}^{-1}$ related to total CB cycling time $f_{m}^{-1}%
+g_{m}^{-1}$, i.e., $\beta=1-n_{0}$) studied by Mehta and Herzog (2008) is
employed. Moreover, the relation $|V_{m2}^{(\operatorname{mod})}%
|=f_{m2}^{(\exp)}x_{m2}^{(\operatorname{mod})}/[1-\beta_{2}^{(\exp)}]$ derived
from Eq. (20) provides a crude model estimate for the CB domain $x_{m2}%
^{(\operatorname{mod})}\thickapprox1.2$ $d$, if their characteristic
attachment time $[f_{2}^{(\exp)}]^{-1}=$ $0.167$ $s$ is also employed. The
rates (shown by \emph{dashed lines}) are determined by the ratio
$f_{m2}^{(\exp)}/g_{m2}^{(\operatorname{mod})}=13$, corresponding to the model
estimate $[g_{m2}^{(\operatorname{mod})}]^{-1}=$ $2.2$ $s$. The \emph{dotted
line} shows a proportion for the assumed transient CB state schematically
drawn for $V/V_{m2}^{(\operatorname{mod})}=0.1$.\newpage

\end{document}